\documentclass[times,twocolumn,final]{elsarticle}

\usepackage{cnf}
\usepackage{graphicx}
\usepackage{amsmath}
\usepackage{amssymb}
\usepackage{bm}
\usepackage{booktabs}
\usepackage{siunitx}
\usepackage{xcolor}
\usepackage{hyperref}
\hypersetup{hidelinks}

\journal{Combustion and Flame}

\begin{document}

\verso{Kalathoor and Oefelein}

\begin{frontmatter}
\title{Event-level compression--chemistry coupling in a supersonic reacting temporal mixing layer}

\author[1]{Sriram P. Kalathoor\corref{cor1}}
\cortext[cor1]{Corresponding author: Daniel Guggenheim School of Aerospace Engineering, Georgia Institute of Technology, Atlanta GA 30332.}
\emailauthor{sriram@gatech.edu}{Sriram P. Kalathoor}
\emailauthor{joseph.oefelein@gatech.edu}{Joseph C. Oefelein}
\author[1]{Joseph C. Oefelein}

\address[1]{Daniel Guggenheim School of Aerospace Engineering, Georgia Institute of Technology, Atlanta GA 30332}

\begin{abstract}
Compression and heat release interact intermittently in high-speed reacting shear layers, and whole-field averages can obscure their coupling. We examine this coupling in a supersonic reacting hydrogen--air temporal mixing layer using time-resolved mid-plane slices from a three-dimensional direct numerical simulation. A fixed dilatation threshold identifies connected compression events, and exothermic heat release and mixture-fraction-gradient activity are then conditioned on the evolving event population. The record separates into startup ($t^*<5$), transition ($5\le t^*<20$), and developed ($t^*\ge 20$) regimes, with the developed regime carrying the persistent compression--chemistry interaction. In this regime, compression appears as a population of intermittent events with no single structure dominating the field. Stronger exothermic response is associated with larger maximum-event area, larger event count, greater compression--heat-release overlap, and smaller distance from compression to the most exothermic regions. Scalar-gradient amplification peaks near zero lag relative to compression-area excursions, whereas the strongest exothermic response precedes peak compression coverage by $\Delta t^*\approx -0.85$. These results show that compression organizes chemistry most clearly through event population, overlap, proximity, and lag, providing an event-level description of compression--chemistry coupling in an open supersonic reacting shear layer.
\end{abstract}


\end{frontmatter}

\section*{Novelty and significance statement}
Supersonic reacting shear layers contain compressive motions, scalar-gradient amplification, and localized heat release in the same evolving field, yet their coupling is often reduced to averaged correlations. The novelty of this study is an event-level characterization of compression--chemistry interaction in which connected compression regions are related to reactive overlap, proximity, morphology, and lag. The significance is that the strongest heat-release response is shown to depend on the placement and timing of the compression population relative to the reactive subset of the layer, providing a compact physical basis for comparing reacting high-speed shear flows.

\section{Introduction}
Compression and dilatation are central to compressible reacting flows, yet their local coupling to chemistry remains difficult to quantify in open shear layers, where intermittent compressive structures coexist with evolving scalar gradients and distributed heat release. In high-speed reacting turbulence, the key issue is how compression is organized into spatial events, how those events intersect the reacting layer, and how the strongest exothermic response is timed relative to compression-area excursions.

The difficulty is that compression is both a kinematic and thermochemical organizer. Kinematically, negative dilatation and the associated strain field can sharpen scalar interfaces and alter the local distribution of mixture-fraction gradients. Thermochemically, heat release depends on the mixture state, local temperature, and reaction progress already present in the compressed region. These two roles need not peak at the same location or time. A compression structure can lie on a scalar interface with strong $|\nabla Z|^2$ while sampling only moderate heat release, or it can pass near a more exothermic region without occupying it directly. This distinction is especially important in open high-speed shear layers, where coherent large-scale motion, small-scale scalar-gradient amplification, and localized reactive activity coexist in the same evolving layer.

This coupling is also spatially intermittent. A plane-averaged correlation can show whether dilatation and heat release vary together, yet it suppresses the geometry of the regions that carry the response. For reacting shear layers, that geometry matters because compression can appear as many small regions, a few larger connected structures, or a fragmented boundary wrapped around scalar interfaces. These arrangements can have similar mean dilatation while placing compression at different distances from the reactive subset of the layer. An event-level description keeps this spatial support visible, which allows the chemistry response to be interpreted through contact, proximity, and timing alongside whole-field moments.

The broader compressible-mixing-layer literature has established several important ingredients for this problem. Experiments and simulations of non-reacting compressible shear layers showed long ago that increasing convective Mach number suppresses layer growth and alters large-scale organization. The experiments of Papamoschou and Roshko \cite{PapamoschouRoshko1988} established the basic growth-rate suppression trend, while the DNS of Vreman \textit{et al.} \cite{VremanSandhamLuo1996} quantified the associated changes in turbulence structure. The high-speed shear-layer study of Pantano and Sarkar \cite{PantanoSarkar2002} and the homogeneous-shear analysis of Blaisdell \textit{et al.} \cite{BlaisdellMansourReynolds1993} further clarified how compressibility modifies the flow structure and dilatational content, and Atoufi \textit{et al.} \cite{AtoufiFathaliLessani2015} examined the associated turbulent-kinetic-energy exchange in temporal mixing layers. Collectively, these studies identified the role of compressibility in reducing spreading and modifying turbulence structure, while leaving open how compressive motions couple to chemistry once heat release and multicomponent mixing become active.

The reacting-shear-layer studies, in turn, showed that heat release feeds back on entrainment, growth, scalar mixing, and reaction-zone structure. Experiments by Hermanson and Dimotakis \cite{HermansonDimotakis1989} found that heat release slightly reduced growth while substantially reducing entrainment. Early numerical work by Givi \textit{et al.} \citep{GiviEtAl1991} showed that heat release can have a stage-dependent effect on temporally developing high-speed reacting layers, with some early enhancement of mixing followed by inhibition at later times. DNS studies by Pantano \textit{et al.} \cite{PantanoSarkarWilliams2003} further demonstrated that heat release modifies the scalar PDF and scalar-dissipation statistics in ways that can reduce overall reaction rate. In reacting turbulent shear flow, Livescu \textit{et al.} \cite{LivescuJaberiMadnia2002} showed that heat release also changes the energy exchange between solenoidal and dilatational motions. More recent three-dimensional DNS of high-speed H$_2$/air mixing layers by Martinez-Ferrer \textit{et al.} \cite{MartinezFerrerEtAl2017a} documented the influence of compressibility and heat release on growth rates and turbulence characteristics, while Martinez-Ferrer \textit{et al.} \cite{MartinezFerrerEtAl2017b} examined the resulting stabilization-zone structure and the location of thermal runaway.

What remains less developed is an event-level description of compression--chemistry coupling itself. Prior work has established the importance of compressibility, heat release, and scalar mixing. Their interaction is often summarized through growth rates, averaged turbulence statistics, scalar PDFs, or ignition/stabilization measures. An event-level view can instead ask whether the strongest exothermic response is synchronized with peak compression, whether compression is linked to chemistry mainly through overlap or through proximity, and how the event population changes as the reacting layer develops.

Several studies come closer to the present question. In reacting turbulent shear flow, Livescu \textit{et al.} \cite{LivescuJaberiMadnia2002} showed from DNS that heat release modifies the explicit dilatational terms, including pressure--dilatation and dilatational dissipation, and can enhance the dilatational kinetic-energy content even while reducing the solenoidal component. In a different configuration, Wagner \textit{et al.} \cite{WagnerEtAl2016} examined the use of planar dilatation as a marker for heat release and found a useful correlation between the two. Those studies support the idea that dilatation and reactive activity are dynamically coupled, and motivate a fixed-threshold, event-based description of how compression structures organize chemistry in a high-speed reacting mixing layer.

Here we study those questions in a supersonic reacting temporal mixing layer using time-resolved mid-plane slices. Connected regions of negative dilatation define the compression events, and heat release and scalar-gradient activity are conditioned on that evolving event population. The resulting description compares how much compression is present, how it is partitioned into connected structures, how closely it lies to the reactive subset of the layer, and how the chemistry response is timed relative to compression-area excursions.

Two choices are central to the interpretation. The compression mask is tied to a fixed dilatation threshold, allowing the selected area to grow or decay with the flow without prescribing it at each time. The record is also separated into startup, transition, and developed intervals before chemistry rankings are formed. These intervals are specific to the temporal evolution of the present layer; the physical distinction is that the developed interval is the part of the record in which compression events persist long enough to carry statistically meaningful heat-release and scalar-gradient information.

\section{Governing equations and data}
The source data come from a three-dimensional direct numerical simulation of a supersonic reacting hydrogen--air temporal mixing layer. The physical configuration follows O'Brien \textit{et al.} \cite{OBrien2014}, and the numerical implementation is based on the compressible reacting-flow formulation described by Oefelein \cite{Oefelein2006-PAS}. The chemical kinetics are represented by a 9-species, 10-step hydrogen--air mechanism based on the hydrogen oxidation model of {\'O} Conaire \textit{et al.} \cite{OConaire2004}. The present analysis uses time-resolved fields on the mid-span plane and focuses on the in-plane organization of velocity, mixture fraction, dilatation, heat release, and scalar-gradient activity. The event statistics are planar quantities: they measure how the sampled cross-section of the reacting layer organizes compression relative to chemistry.

\begin{table*}[t]
\centering
\footnotesize
\setlength{\tabcolsep}{7pt}
\caption{Physical configuration used for the reacting temporal mixing layer. The setup follows the hydrogen--air configuration of O'Brien \textit{et al.} \cite{OBrien2014}.}
\label{tab:physical_config}
\begin{tabular}{p{0.25\textwidth}p{0.68\textwidth}}
\toprule
Quantity & Value \\
\midrule
Flow configuration & Supersonic reacting H$_2$--air temporal mixing layer \\
Mean pressure & \SI{202.65}{\kilo\pascal} \\
Oxidizer stream & Air, $T=\SI{1500}{\kelvin}$, $M=2.25$, $Z=0$ \\
Fuel stream & H$_2$/N$_2$ mixture, $X_{\mathrm{H}_2}=0.06$, $T=\SI{500}{\kelvin}$, $M=2.73$, $Z=1$ \\
Species composition & $Y_{\mathrm{O}_2}=0.2329$, $Y_{\mathrm{N}_2}=0.7671$ in air; $Y_{\mathrm{H}_2}=0.00457$, $Y_{\mathrm{N}_2}=0.99543$ in the fuel stream \\
\bottomrule
\end{tabular}
\end{table*}

The mid-plane record is well suited to this objective because it resolves event birth, growth, overlap, and displacement across a long temporal history. The statistics should be read as planar organization measures. Three-dimensional structure enters through the source flow and physical configuration; the event analysis asks how one resolved section of that flow arranges compression, scalar gradients, and heat release.

The stream conditions in table \ref{tab:physical_config} place a hot oxidizer stream against a colder dilute hydrogen stream at supersonic speeds. The resulting layer combines strong compressibility, large temperature contrast, and finite-rate chemistry in the same evolving shear region. In that setting, compression can act through several routes at once: it can alter local density and temperature, sharpen mixture-fraction gradients, and change the residence of reactive mixture in regions of strong strain. The event analysis in this work is designed to separate these effects at the level of spatial organization.

\section{Compression mask and conditioning}
We define compression using the dilatation field $\nabla\cdot\bm{u}$. The dilatation is the full three-dimensional velocity-gradient trace,
\begin{equation}
\nabla\cdot\bm{u}
=\frac{\partial u}{\partial x}
+\frac{\partial v}{\partial y}
+\frac{\partial w}{\partial z},
\end{equation}
evaluated on the mid-span plane. The planar restriction enters through the connected-component, area, overlap, and conditioning statistics. A fixed threshold is set from the initial field to avoid trivial constant area fractions:
\begin{equation}
\begin{aligned}
\Omega_c(t) &= \{\bm{x}: \nabla\cdot\bm{u}(\bm{x},t) \le \eta_c\},\\
\eta_c &= P_{p_c}[\nabla\cdot\bm{u}(\bm{x},t_{\mathrm{init}})],
\end{aligned}
\end{equation}
where $P_{p_c}$ denotes the $p_c$-percentile (here $p_c=5$) at the initial time $t_{\mathrm{init}}$. Expansion is defined analogously by
\begin{equation}
\begin{aligned}
\Omega_e(t) &= \{\bm{x}: \nabla\cdot\bm{u}(\bm{x},t) \ge \eta_e\},\\
\eta_e &= P_{p_e}[\nabla\cdot\bm{u}(\bm{x},t_{\mathrm{init}})],
\end{aligned}
\end{equation}
with $p_e=95$. The corresponding area fractions are
\begin{equation}
\alpha_c(t)=\frac{|\Omega_c(t)|}{|\Omega|}, \qquad
\alpha_e(t)=\frac{|\Omega_e(t)|}{|\Omega|},
\end{equation}
where $\Omega$ denotes the full slice domain. Given any field $\phi$, we compute compression-conditioned means
\begin{equation}
\langle \phi \rangle_{c}(t) = \frac{1}{|\Omega_c(t)|}\int_{\Omega_c(t)} \phi(\bm{x},t)\, dA,
\end{equation}
and unconditional plane means
\begin{equation}
\langle \phi \rangle_{\Omega}(t) = \frac{1}{|\Omega|}\int_{\Omega} \phi(\bm{x},t)\, dA.
\end{equation}
Heat release rate is obtained from species enthalpies and net reaction rates,
\begin{equation}
\dot{q} = -\sum_k h_k \dot{\omega}_k.
\end{equation}
With this sign convention, stronger exothermic activity corresponds to more negative $\dot{q}$, so exothermic ``high-activity'' masks are taken from the lower tail of the $\dot{q}$ distribution. At each time, the exothermic and high-gradient indicator sets are
\begin{equation}
\begin{aligned}
\Omega_q(t) &= \{\bm{x}: \dot{q}(\bm{x},t)\le P_{p_q}[\dot{q}(\bm{x},t)]\},\\
\Omega_g(t) &= \{\bm{x}: |\nabla Z|^2(\bm{x},t)\ge P_{p_g}[|\nabla Z|^2(\bm{x},t)]\},
\end{aligned}
\end{equation}
with $p_q=5$ and $p_g=95$. We denote the mixture fraction gradient magnitude by $|\nabla Z|^2$. For consistency, we use the nondimensionalization of the form
\begin{equation}
t^* = \frac{t U_{c,0}}{\delta_{\omega,0}}, \qquad
(\nabla\cdot\bm{u})^* = (\nabla\cdot\bm{u})\frac{\delta_{\omega,0}}{U_{c,0}},
\end{equation}
\begin{equation}
|\nabla Z|^{2*} = |\nabla Z|^2 \delta_{\omega,0}^2, \qquad
\dot{q}^* = \frac{\dot{q}}{\dot{q}_{\mathrm{ref}}},
\end{equation}
where $\dot{q}_{\mathrm{ref}}$ is a fixed representative exothermic heat-release magnitude for the present configuration. Event area is normalized by $\delta_{\omega,0}^2$, distances by $\delta_{\omega,0}$, and event lifetimes are reported as $\tau^*=\tau U_{c,0}/\delta_{\omega,0}$.

The fixed threshold is central to the interpretation. A threshold recomputed from each field would enforce a prescribed compression area at every time and would remove the area growth that marks the transition from sparse compression to a developed event population. With the threshold held fixed, the measured compression fraction becomes a flow response. Increased overlap with exothermic or high-gradient regions then reflects a real change in the spatial relation between compression and chemistry, and a changing event count reflects how the layer partitions compression into connected structures. During startup this criterion selects very little area, marking the part of the record before compression has developed into a meaningful chemistry-conditioned population.

This choice also separates the support of compression from the response of the chemistry. The compression mask identifies where the flow is locally compressive; it does not assume that those regions are reactive. Heat release and scalar-gradient activity are evaluated after the compression support has been selected, so the conditional statistics measure how the reacting layer is sampled by compression. This distinction is important in a supersonic reacting mixing layer because strong scalar gradients can be generated by kinematic sharpening even when the local mixture state is not the most exothermic one. The overlap and distance measures separate direct co-location from nearby organization along the same scalar layer.

The conditioning has a direct physical interpretation. Negative dilatation identifies local volumetric compression, while the mixture-fraction-gradient field identifies regions where molecular mixing and scalar-interface sharpening are active. Heat release adds an additional selectivity because exothermic activity depends on both the local mixture state and the thermochemical progress of the reacting layer. A compression event can influence the sampled chemistry through three related pathways. It can intersect the exothermic set directly, it can lie close enough to that set for compression and reaction to occupy neighboring parts of the same scalar layer, and it can increase the local scalar-gradient activity that later feeds the reaction-zone structure. The overlap, distance, and lag measures below are designed to keep these pathways distinct. Direct overlap measures co-location, distance measures nearby organization without requiring mask intersection, and lag measures whether scalar-gradient and heat-release responses are synchronized with the growth of compression area. The event-level formulation separates these pathways in the sampled plane.

\section{Results and discussion}
The time record separates into three case-specific phases. A startup regime ($t^*<5$) contains little resolved compression activity and is dominated by the initial establishment of the compression support. A transition regime ($5\le t^*<20$) brings rapid growth of scalar-gradient activity and the first sustained compression events. The developed regime ($t^*\ge 20$) carries the statistically relevant compression--chemistry coupling and is the primary focus of the interpretation below.

\begin{figure*}[t]
  \centering
  \includegraphics[width=0.48\textwidth]{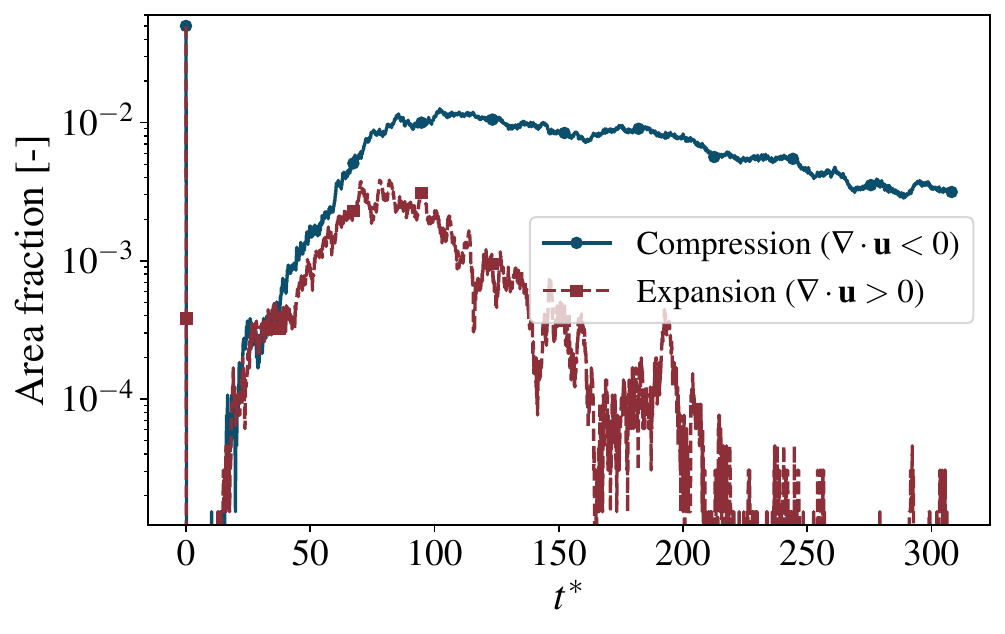}
  \includegraphics[width=0.48\textwidth]{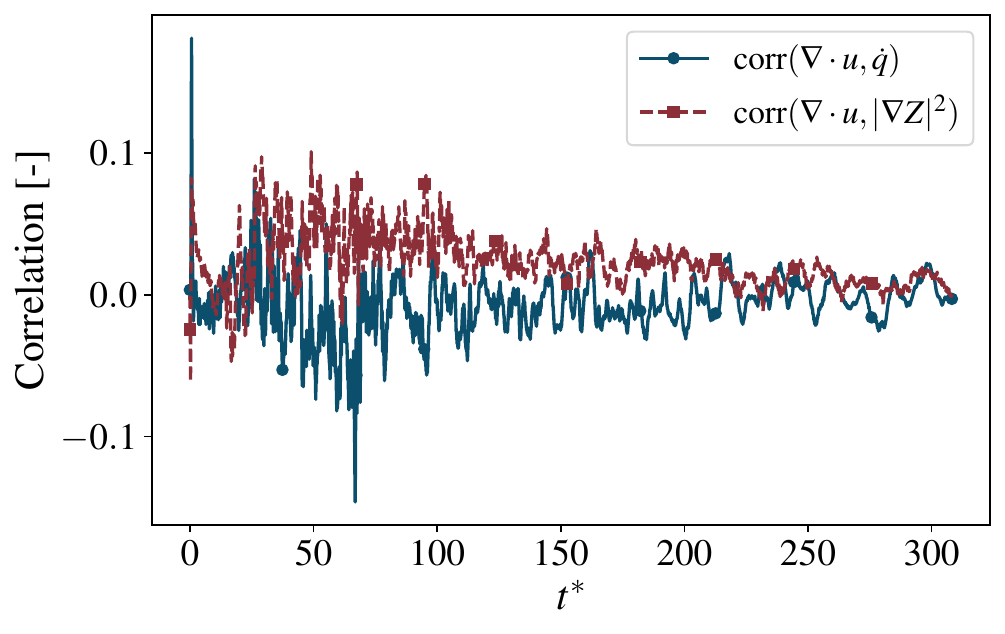}
\caption{Baseline temporal metrics in $t^*$: compression/expansion area fractions (left, shown as deviations when nearly constant) and correlations between dilatation and chemistry indicators (right).}\label{fig:baseline_time}
\end{figure*}

\begin{figure*}[t]
  \centering
  \includegraphics[width=0.92\textwidth]{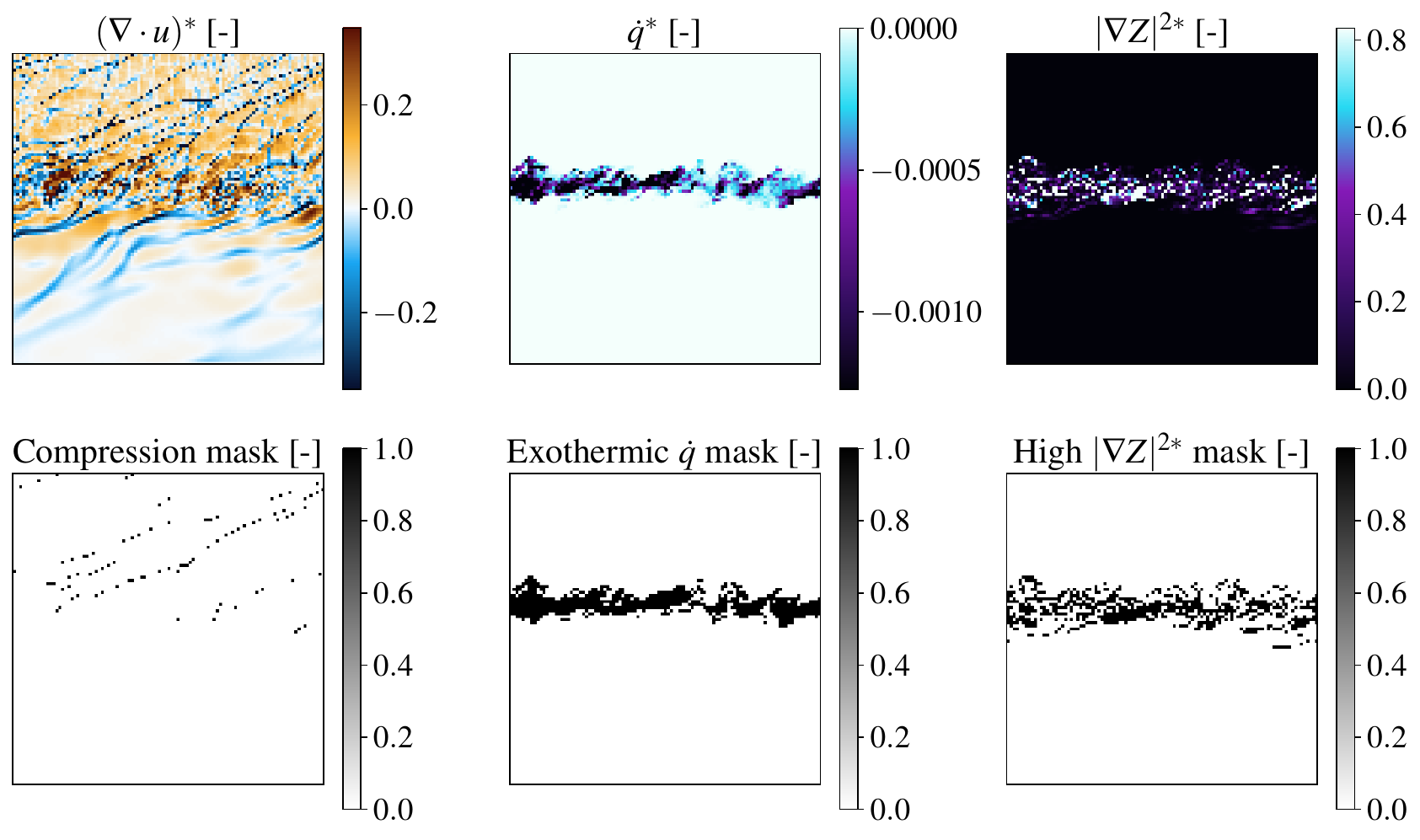}
\caption{Representative developed-regime field structure. The panels show $(\nabla\cdot\bm{u})^*$, $\dot{q}^*$, $|\nabla Z|^{2*}$, and the associated compression, exothermic, and high-gradient masks used in the event, overlap, and proximity measures.}\label{fig:structure}
\end{figure*}

\begin{figure*}[t]
  \centering
  \includegraphics[width=0.48\textwidth]{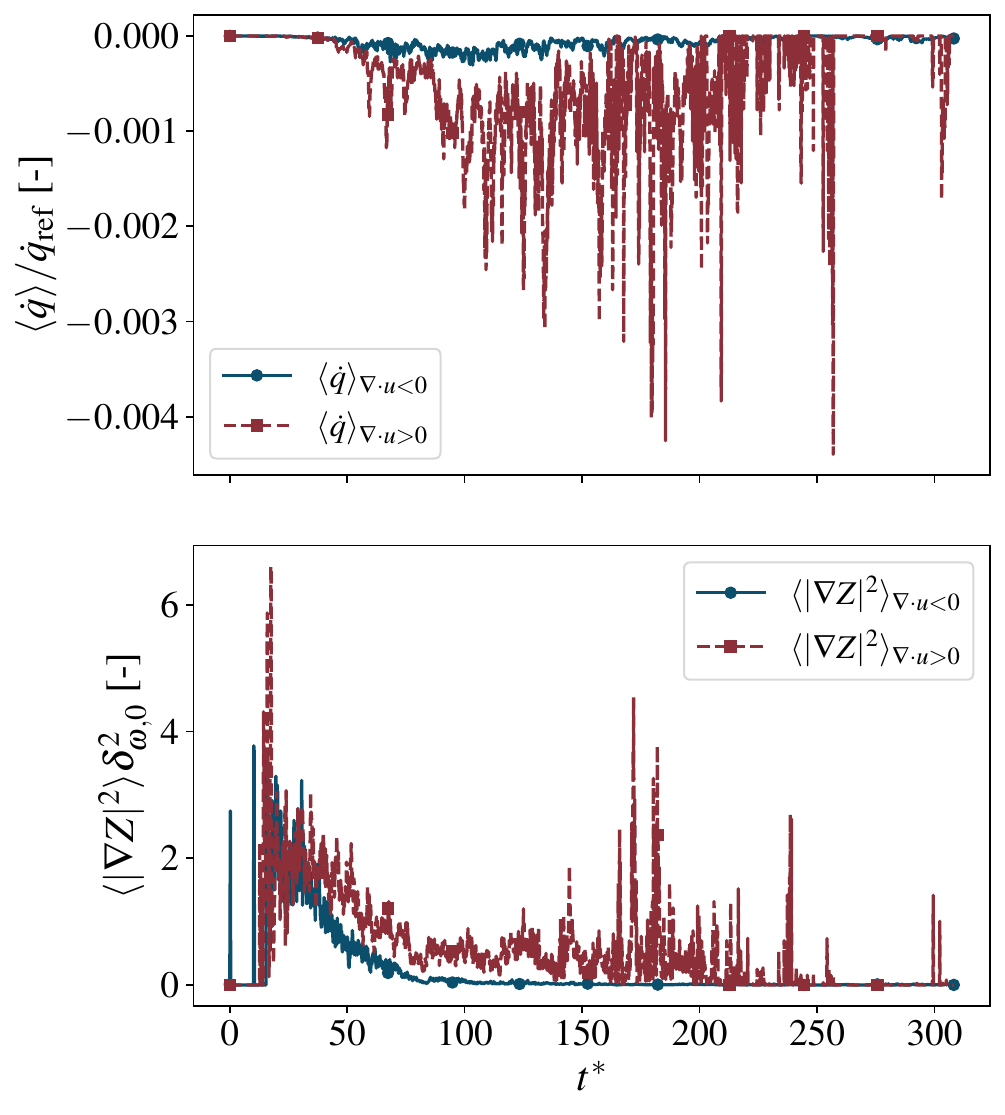}
  \includegraphics[width=0.48\textwidth]{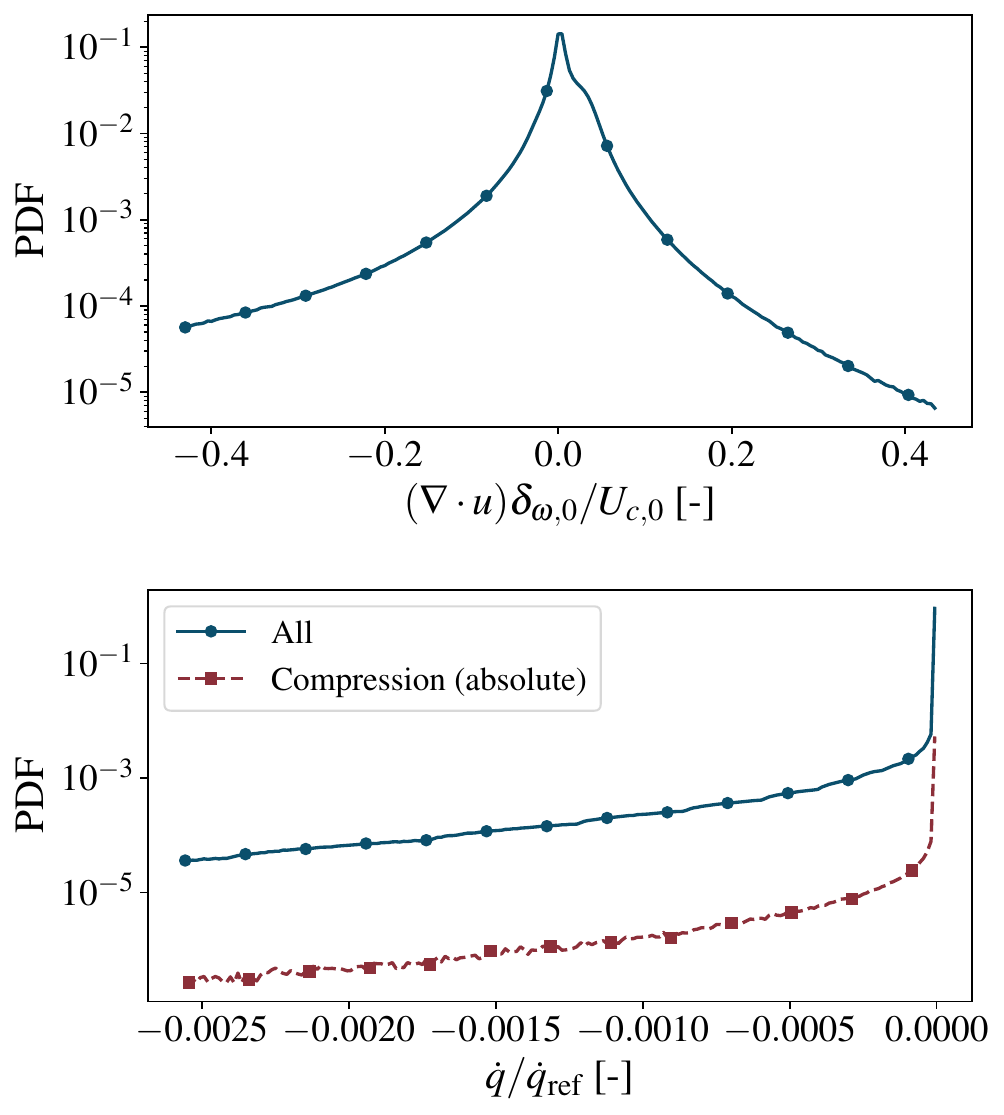}
\caption{Conditional statistics and PDFs. The left panel shows $\langle \dot{q}^* \rangle_c$ and $\langle |\nabla Z|^{2*} \rangle_c$, while the right panel shows the PDFs of $(\nabla\cdot\bm{u})^*$ and $\dot{q}^*$. The compression PDF is reported as an absolute probability by weighting the conditional histogram by the instantaneous compression fraction $\alpha_c(t)$.}\label{fig:baseline_stats}
\end{figure*}

\subsection{Regime decomposition}
The first measures are the compression and expansion area fractions,
\begin{equation}
\alpha_c(t)=\frac{|\Omega_c(t)|}{|\Omega|}, \qquad
\alpha_e(t)=\frac{|\Omega_e(t)|}{|\Omega|},
\end{equation}
together with full-plane correlations between dilatation and chemistry-related fields,
\begin{equation}
\rho_{f,g}(t)=\frac{\langle (f-\langle f\rangle_{\Omega})(g-\langle g\rangle_{\Omega})\rangle_{\Omega}}
{\sigma_f \sigma_g}.
\end{equation}
\noindent
The use of Pearson correlation follows the classical definition \cite{Pearson1895}. These metrics show whether the fixed-threshold compression set occupies a time-varying portion of the plane and whether that occupied area evolves with reactive activity. Figure \ref{fig:baseline_time} shows a strong evolution of the global compression area fraction. The startup interval contains little compression area and almost no conditional heat-release response. During transition, conditioned $|\nabla Z|^{2*}$ grows rapidly and organized compression structures first appear. The developed interval sustains nonzero compression area and contains the statistically meaningful event population. This regime split identifies the part of the record that carries persistent compression--chemistry coupling. The full record establishes the evolution of the layer, while the main physical conclusions are drawn from the developed regime. The specific values of the boundaries are tied to this record; the underlying physical division is between sparse initial compression, emerging compression--scalar-gradient interaction, and sustained compression--chemistry coupling. Because the compression threshold is fixed, the occupied area is itself a dynamical observable. That interpretation is broadly consistent with prior compressible mixing-layer studies in which the effective growth and organization of the layer evolve substantially with compressibility level without remaining tied to a fixed fraction of the domain \cite{PapamoschouRoshko1988,VremanSandhamLuo1996,PantanoSarkar2002}.

\begin{table*}[t]
\centering
\footnotesize
\setlength{\tabcolsep}{5pt}
\caption{Regime-averaged event measures. Here $\alpha_c$ is the compression area fraction, $N_c$ is the number of connected compression events, $\chi_q$ and $\chi_g$ are the compression overlap fractions with exothermic heat release and high $|\nabla Z|^{2*}$, and $\bar d_q$ and $\bar d_g$ are mean distances from compression to those two activity sets. Startup distances are omitted because the compression set is nearly empty and exothermic overlap vanishes.}
\label{tab:regime_summary}
\begin{tabular}{lccccccc}
\toprule
Regime & $t^*$ interval & $\langle\alpha_c\rangle$ & $\langle N_c\rangle$ &
$\langle\chi_q\rangle$ & $\langle\chi_g\rangle$ &
$\langle\bar d_q\rangle/\delta_{\omega,0}$ &
$\langle\bar d_g\rangle/\delta_{\omega,0}$ \\
\midrule
Startup & $t^*<5$ & $5.87\times10^{-4}$ & 4.42 & 0 & $4.13\times10^{-5}$ & -- & -- \\
Transition & $5\le t^*<20$ & $4.53\times10^{-4}$ & 4.38 & 0.006 & 0.561 & 1.13 & 0.33 \\
Developed & $t^*\ge20$ & 0.00722 & 63.27 & 0.054 & 0.073 & 5.37 & 4.96 \\
\bottomrule
\end{tabular}
\end{table*}

The regime averages in table \ref{tab:regime_summary} show the same separation. During transition, compression overlaps strongly with high scalar-gradient activity, while exothermic overlap remains small. Compression is already organized near the scalar layer, but the most selective heat-release response is still developing. In the developed interval, the compression area fraction and event count increase substantially, and both overlap measures remain finite over a sustained record. The mean distances also increase as the reacting layer spreads across a broader cross-stream region; the more useful comparison is the relative proximity of compression to the two chemistry-related sets within the same regime. In the developed regime those distances are comparable, while the overlap fractions show that high-gradient regions are sampled more frequently than the strongest exothermic regions. Scalar-gradient amplification is spatially broader, and the strongest heat release occupies a more selective subset of the layer.

\subsection{Baseline conditioned response}
The baseline response is seen first in the full-field and compression-conditioned distributions. For any field $\phi$, the compression-conditioned mean is
\begin{equation}
\langle \phi \rangle_{c}(t) = \frac{1}{|\Omega_c(t)|}\int_{\Omega_c(t)} \phi(\bm{x},t)\, dA,
\end{equation}
and the corresponding unconditional mean is taken over the full plane. In addition, PDFs of $(\nabla\cdot\bm{u})^*$ and $\dot{q}^*$ are formed from plane histograms, with the compression-conditioned PDF reported as an absolute probability by weighting the conditional histogram with $\alpha_c(t)$. Figure \ref{fig:baseline_stats} confirms that compression preferentially samples regions of stronger scalar gradients and more exothermic chemistry. Because the heat-release convention used here makes exothermic activity more negative, stronger coupling appears as more negative conditional $\langle\dot{q}^*\rangle_c$ values. The strongest correlations are intermittent and regime-dependent, with the transition interval carrying the sharpest rise in $|\nabla Z|^{2*}$. These intermittent signatures motivate an event-based view. Their amplitude is controlled by the geometry and organization of the compression population as much as by record-wide mean behavior. Figure \ref{fig:structure} gives a representative field-level view of that organization, with compression masks, exothermic masks, and high-gradient masks occupying related yet distinct spatial support. The conditional means and PDFs also separate the two chemistry-related scalars: $|\nabla Z|^{2*}$ responds more immediately to compression, while $\dot{q}^*$ carries a more delayed and spatially selective signature. This behavior is compatible with the broader reacting-shear-layer literature, which has emphasized that heat release modifies scalar mixing and reaction-zone structure in a spatially selective manner \cite{HermansonDimotakis1989,PantanoSarkarWilliams2003, MartinezFerrerEtAl2017a,MartinezFerrerEtAl2017b}.

\subsection{Event population and overlap}
\begin{figure*}[t]
  \centering
  \includegraphics[width=0.48\textwidth]{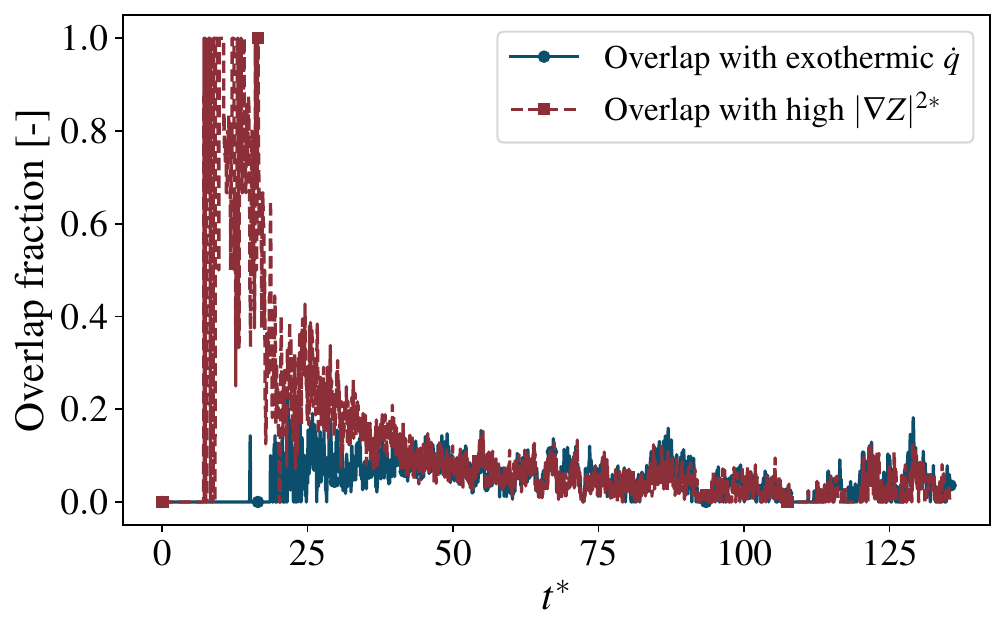}
  \includegraphics[width=0.48\textwidth]{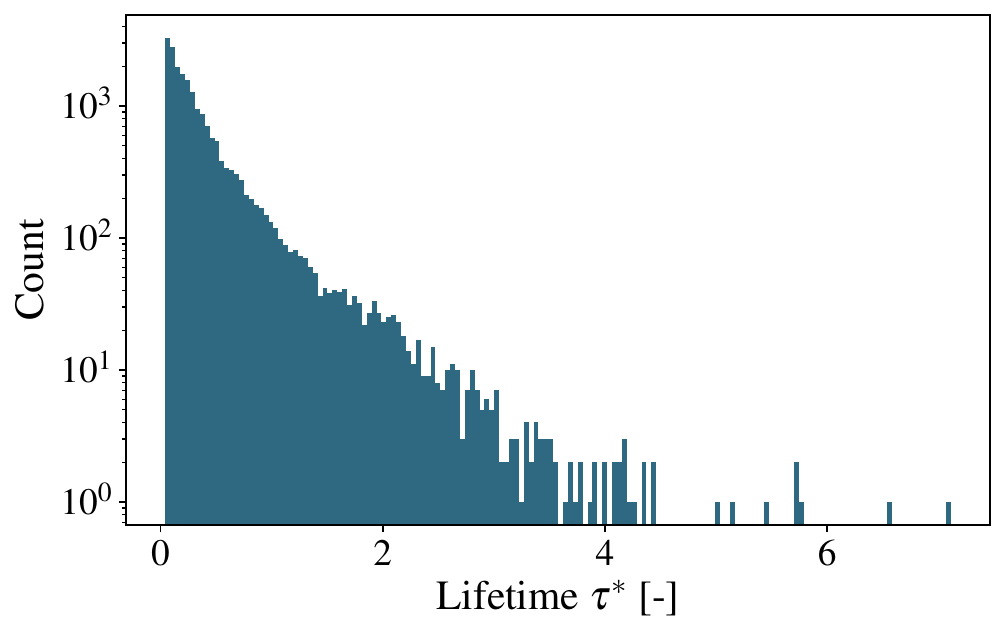}
\caption{Compression event overlap and persistence: overlap fractions $\chi_q$ and $\chi_g$ of compression with exothermic $\dot{q}$ and high $|\nabla Z|^{2*}$ regions (left), and event lifetimes $\tau^*$ (right).}
  \label{fig:events_overlap_life}
\end{figure*}

To resolve that population, compression events are defined as connected components of $\Omega_c(t)$ under 8-neighbour connectivity,
\begin{equation}
\Omega_c(t)=\bigcup_{j=1}^{N_c(t)} \mathcal{C}_j(t),
\end{equation}
\noindent
using standard connected-component labelling ideas \cite{RosenfeldPfaltz1966}. The corresponding event count is $N_c(t)$, and the event areas are
\begin{equation}
\begin{aligned}
A_j(t)&=\int_{\mathcal{C}_j(t)} dA,\\
\bar{A}(t)&=\frac{1}{N_c(t)}\sum_{j=1}^{N_c(t)} A_j(t),\\
A_{\max}(t)&=\max_j A_j(t).
\end{aligned}
\end{equation}
Overlap with exothermic and high-gradient activity is quantified by
\begin{equation}
\chi_q(t)=\frac{|\Omega_c(t)\cap \Omega_q(t)|}{|\Omega_c(t)|}, \qquad
\chi_g(t)=\frac{|\Omega_c(t)\cap \Omega_g(t)|}{|\Omega_c(t)|}.
\end{equation}
\noindent
These overlap measures are directly related to classical set-similarity ideas \cite{Jaccard1901}.
\begin{figure}[t]
  \centering
  \includegraphics[width=\columnwidth]{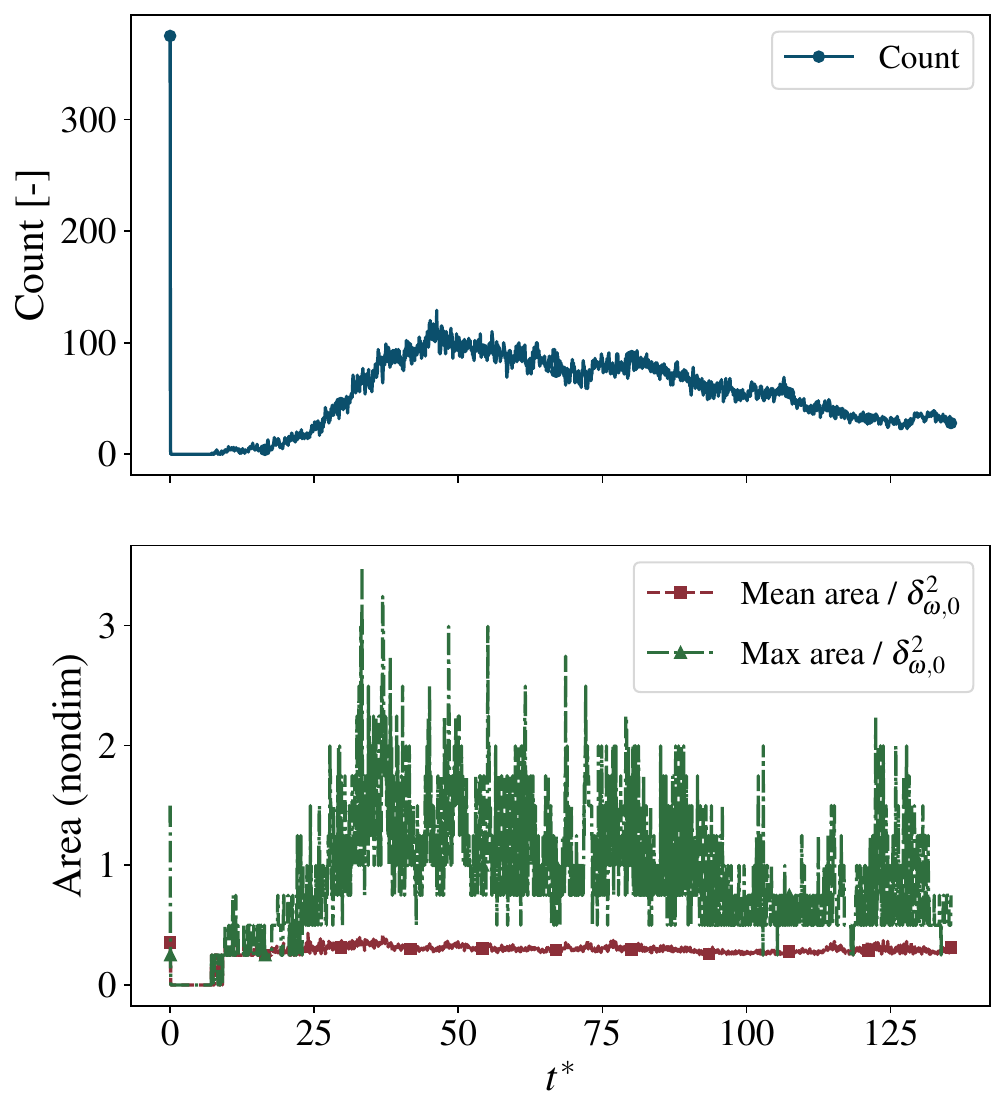}
\caption{Compression event population statistics: event count and event area $A^*$ across the record. The upper panel reports the number of connected compression events, while the lower panel reports the mean and maximum event areas.}\label{fig:events_population}
\end{figure}
The event-population statistics in figure \ref{fig:events_population} show that developed-regime compression consists of many connected structures with modest mean area and an intermittent large-event tail. Lifetime statistics in figure \ref{fig:events_overlap_life} are obtained by associating components across successive fields through maximum overlap. Many events are short-lived, with a non-negligible tail extending to longer nondimensional lifetimes. The chemistry signal is tied to the evolving ensemble of compression events.

Overlap statistics in figure \ref{fig:events_overlap_life} sharpen the chemistry picture. Compression overlaps more consistently with high-$|\nabla Z|^{2*}$ than with the strongest exothermic heat-release regions, which indicates that scalar-gradient amplification is the more immediate signature of compression. Even so, exothermic overlap is systematic in the developed regime and grows with event organization: larger maximum-event area, higher event count, and larger exothermic-overlap fraction all correspond to more negative conditional $\langle\dot{q}^*\rangle_c$. In the developed regime, quartile-based summaries show a nearly monotone progression: time instants in the highest exothermic-overlap quartile are markedly more exothermic than those in the lowest quartile, and similar ordering appears when the data are grouped by event count or maximum-event area. Compression organization carries chemistry information because the reactive response depends on how compression is partitioned into events, how large those events become, and how strongly they intersect the exothermic subset of the heat-release field. This interpretation is consistent with the full-field view in figure \ref{fig:structure}. The exothermic set is more spatially selective than the high-gradient set, so event organization becomes especially important for locating the strongest negative $\dot{q}^*$ samples. The result also complements prior observations that heat release suppresses entrainment and alters growth in reacting shear layers \cite{HermansonDimotakis1989,GiviEtAl1991}. Here that global suppression is resolved into a more local statement: chemistry responds most strongly when compression is organized into sufficiently large and numerous events that also intersect the exothermic subset of the layer.

\begin{table*}[t]
\centering
\footnotesize
\setlength{\tabcolsep}{6pt}
\caption{Developed-regime ranking of compression-event measures against the compression-conditioned heat release. Correlations are computed with $\langle\dot{q}^*\rangle_c$; more negative response values correspond to stronger exothermic activity. Quartiles are ordered by increasing value of the conditioning measure, and the response columns report $10^4\langle\dot{q}^*\rangle_c$.}
\label{tab:developed_ranking}
\begin{tabular}{lcccl}
\toprule
Conditioning measure & Correlation & Q1 response & Q4 response & Physical reading \\
\midrule
$\chi_q$ & -0.55 & -0.13 & -1.39 & direct compression--heat-release overlap \\
$\bar d_q/\delta_{\omega,0}$ & 0.28 & -1.06 & -0.47 & proximity to exothermic heat release \\
$A_{\max}/\delta_{\omega,0}^2$ & -0.36 & -0.46 & -1.18 & large-event support \\
$N_c$ & -0.43 & -0.44 & -1.43 & distributed event population \\
\bottomrule
\end{tabular}
\end{table*}

The developed-regime ranking in table \ref{tab:developed_ranking} makes the hierarchy quantitative. Exothermic overlap gives the clearest monotone separation: the highest-overlap quartile has an order-of-magnitude stronger compression-conditioned exothermic response than the lowest-overlap quartile on the nondimensional scale used in the figures. Distance to exothermic heat release carries the same physical message with the opposite ordering, since nearer compression regions correspond to more negative $\langle\dot{q}^*\rangle_c$. Event count and maximum event area also rank the response, which indicates that the population-level organization of compression matters in addition to the local geometry of individual regions. The strongest chemistry signature appears when compression is numerous or spatially extended and also lies close to the exothermic subset of the reacting layer.

\begin{table*}[t]
\centering
\footnotesize
\setlength{\tabcolsep}{5pt}
\caption{Threshold sensitivity of developed-regime event rankings. The baseline uses $p_c=5$ for compression and $p_q=p_g=95$ for chemistry-related support. Response separation is the top-minus-bottom quartile change in $10^4\langle\dot{q}^*\rangle_c$ when ranking by $\chi_q$; negative values indicate stronger exothermic response at larger exothermic overlap.}
\label{tab:threshold_sensitivity}
\begin{tabular}{lccccc}
\toprule
Case & $p_c$ & $p_q,p_g$ & $\rho(\chi_q,\langle\dot{q}^*\rangle_c)$ & $\rho(N_c,\langle\dot{q}^*\rangle_c)$ & $\Delta_{\chi_q}(10^4\langle\dot{q}^*\rangle_c)$ \\
\midrule
baseline & 5 & 95 & -0.68 & -0.43 & -1.31 \\
lower compression support & 4 & 95 & -0.67 & -0.38 & -1.29 \\
higher compression support & 6 & 95 & -0.70 & -0.47 & -1.32 \\
broader chemistry support & 5 & 94 & -0.62 & -0.43 & -1.22 \\
narrower chemistry support & 5 & 96 & -0.72 & -0.43 & -1.35 \\
\bottomrule
\end{tabular}
\end{table*}

This ordering is stable across the tested support definitions. In table \ref{tab:threshold_sensitivity}, varying the compression percentile and the chemistry-related activity percentile leaves the sign and relative size of the main correlations unchanged. The exothermic-overlap correlation remains the strongest ranking measure across all tested supports, the event-count correlation remains negative, and the quartile separation in conditioned heat release varies only modestly.

The sensitivity result matters because the masks represent spatial support. A looser compression support includes more moderately compressive material around the event cores, while a tighter support emphasizes the strongest local compression. Broadening or narrowing the chemistry support changes how much of the exothermic tail is retained. The stable signs in table \ref{tab:threshold_sensitivity} show that the inferred ordering is robust across the tested cutoffs. The dominant signal remains geometric and physical: compression is most predictive of heat release when it occupies, approaches, or repeatedly samples the exothermic subset of the reacting layer.

\subsection{Lead/lag structure and proximity}
\begin{figure}[!b]
  \centering
  \includegraphics[width=0.96\columnwidth]{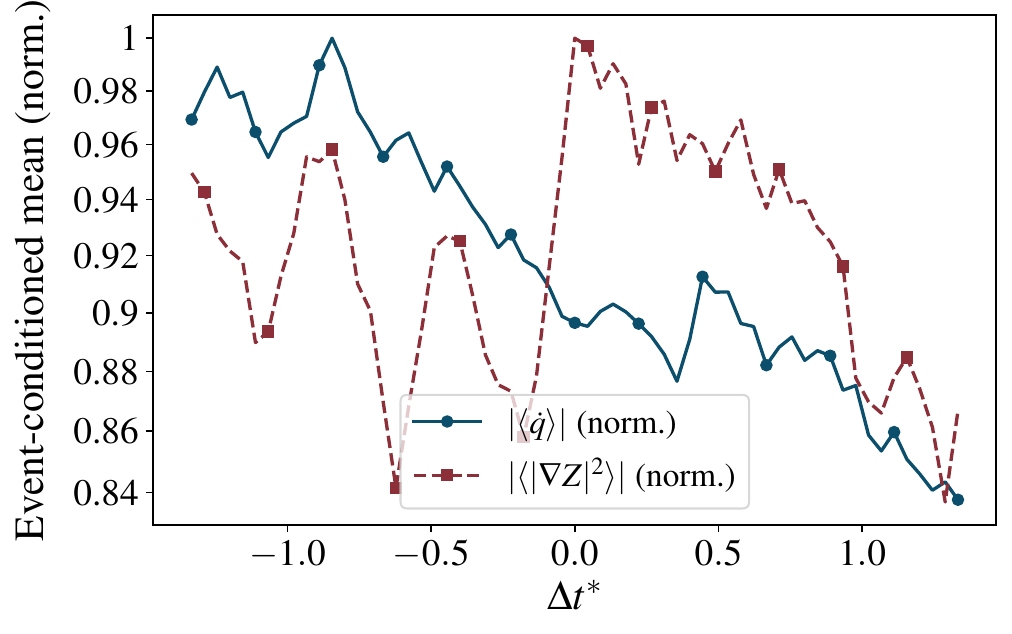}
\caption{Event-triggered responses $\mathcal{R}_{\dot{q}}$ and $\mathcal{R}_{|\nabla Z|^{2}}$ versus lag $\Delta t^*$, shown on a normalized log scale.}\label{fig:events_coupling}
\end{figure}

To quantify timing relative to strong compression activity, we define the set $\mathcal{E}$ of upper-tail compression-area excursions and compute the event-triggered response
\begin{equation}
\mathcal{R}_{\phi}(\ell)=\frac{1}{|\mathcal{E}|}\sum_{t_k\in\mathcal{E}}
\langle \phi \rangle_c(t_k+\ell),
\end{equation}
with $\phi\in\{\dot{q}^*, |\nabla Z|^{2*}\}$. The right panel of figure \ref{fig:events_coupling} shows that the event-conditioned $|\nabla Z|^{2*}$ response peaks near zero lag, whereas the largest-magnitude exothermic $\langle\dot{q}^*\rangle_c$ response precedes the peak compression-area excursion by $\Delta t^*\approx -0.85$. This indicates that the strongest exothermic activity is displaced from the time of maximum compression coverage. Scalar-gradient amplification responds much more nearly synchronously.

\begin{figure}[t]
  \centering
  \includegraphics[width=\columnwidth]{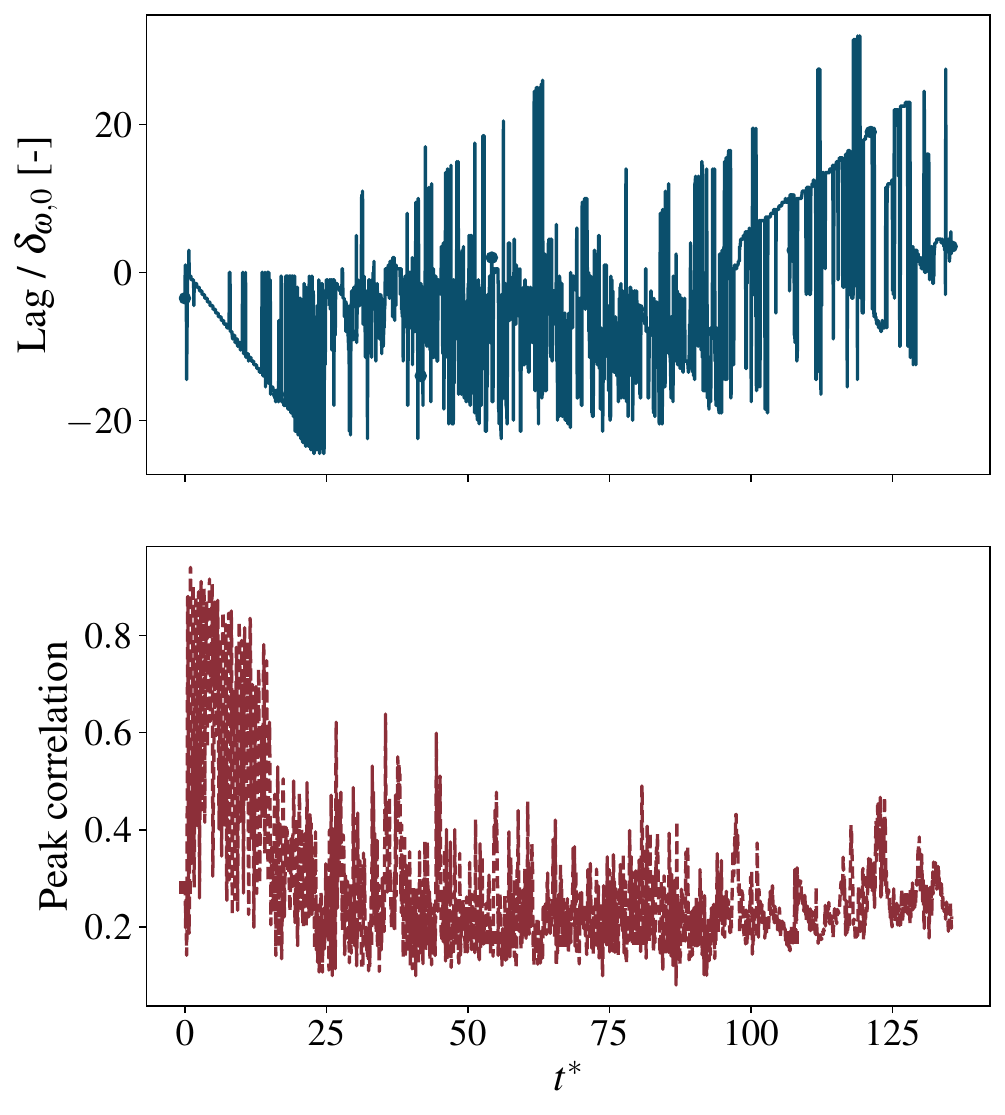}
\caption{Cross-stream lag $\Delta y^*$ between compression and heat release, together with the corresponding peak cross-correlation history.}
  \label{fig:lag}
\end{figure}

\begin{figure*}[t]
  \centering
  \includegraphics[width=0.48\textwidth]{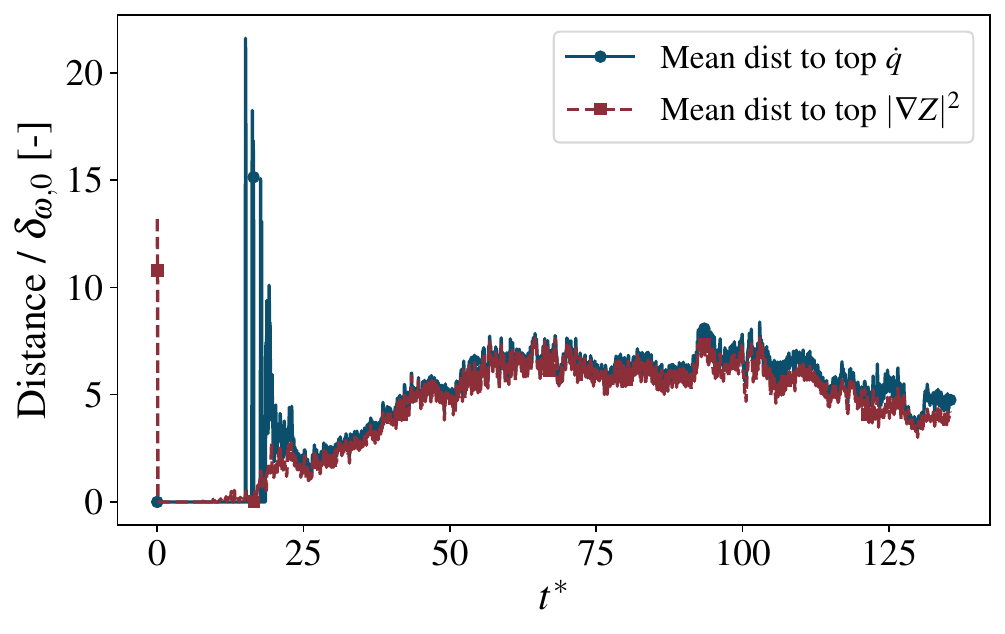}
  \includegraphics[width=0.48\textwidth]{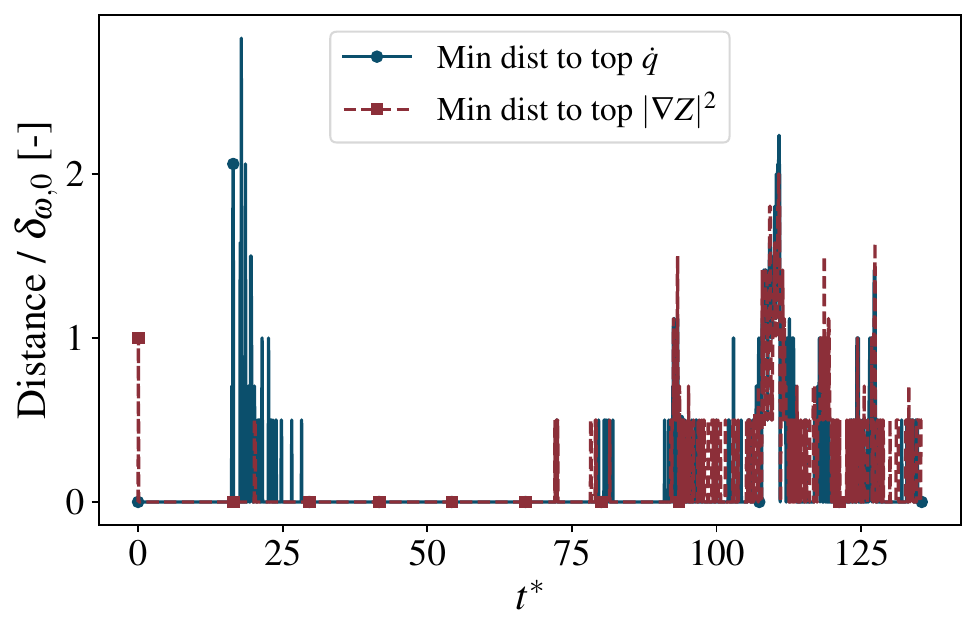}
\caption{Compression-to-chemistry distances: mean and minimum distances from compression regions to exothermic $\dot{q}$ and high $|\nabla Z|^{2*}$ regions, corresponding to $\bar d$ and $d_{\min}$ statistics derived from the distance transforms.}\label{fig:dist_stats}
\end{figure*}

The cross-stream lag and distance measures in figures \ref{fig:lag} and \ref{fig:dist_stats} support the same interpretation. The cross-stream lead/lag is computed from the peak of the streamwise averaged cross-correlation
\begin{equation}
C_{fg}(\ell_y,t)=\sum_y \bar{f}(y,t)\,\bar{g}(y+\ell_y,t),
\end{equation}
between $f=(\nabla\cdot\bm{u})^*$ and $g=\dot{q}^*$, while proximity is measured from Euclidean distance transforms \cite{FelzenszwalbHuttenlocher2012}, with a combustion-context distance-function reference \cite{NilssonBai2003}. If $d_q(\bm{x},t)$ and $d_g(\bm{x},t)$ denote the distances to $\Omega_q(t)$ and $\Omega_g(t)$, respectively, then
\begin{equation}
\begin{aligned}
\bar{d}_q(t)&=\frac{1}{|\Omega_c(t)|}\int_{\Omega_c(t)} d_q(\bm{x},t)\,dA,\\
d_{q,\min}(t)&=\min_{\bm{x}\in\Omega_c(t)} d_q(\bm{x},t),
\end{aligned}
\end{equation}
with analogous definitions for the high-gradient set. Compression structures remain systematically closer to high-$|\nabla Z|^{2*}$ than to the strongest exothermic $\dot{q}$ regions. Time instants at which compression lies closest to exothermic heat release exhibit the most negative conditional $\langle\dot{q}^*\rangle_c$. Proximity to exothermic $\dot{q}$ and direct overlap with exothermic regions provide the clearest chemistry ranking among the developed-regime metrics. Distance remains informative even when instantaneous overlap is small, making it a useful companion to overlap. This emphasis on proximity is consistent with scalar-mixing results for reacting shear layers \cite{PantanoSarkarWilliams2003}, which showed that reaction rates are strongly mediated by local scalar-dissipation structure. The distance-based measures add a geometric view of that relation by indicating how near compression lies to the subset of the flow where the strongest chemistry is expressed.

\subsection{Event geometry and alignment}
\begin{figure}[t]
  \centering
  \includegraphics[width=\columnwidth]{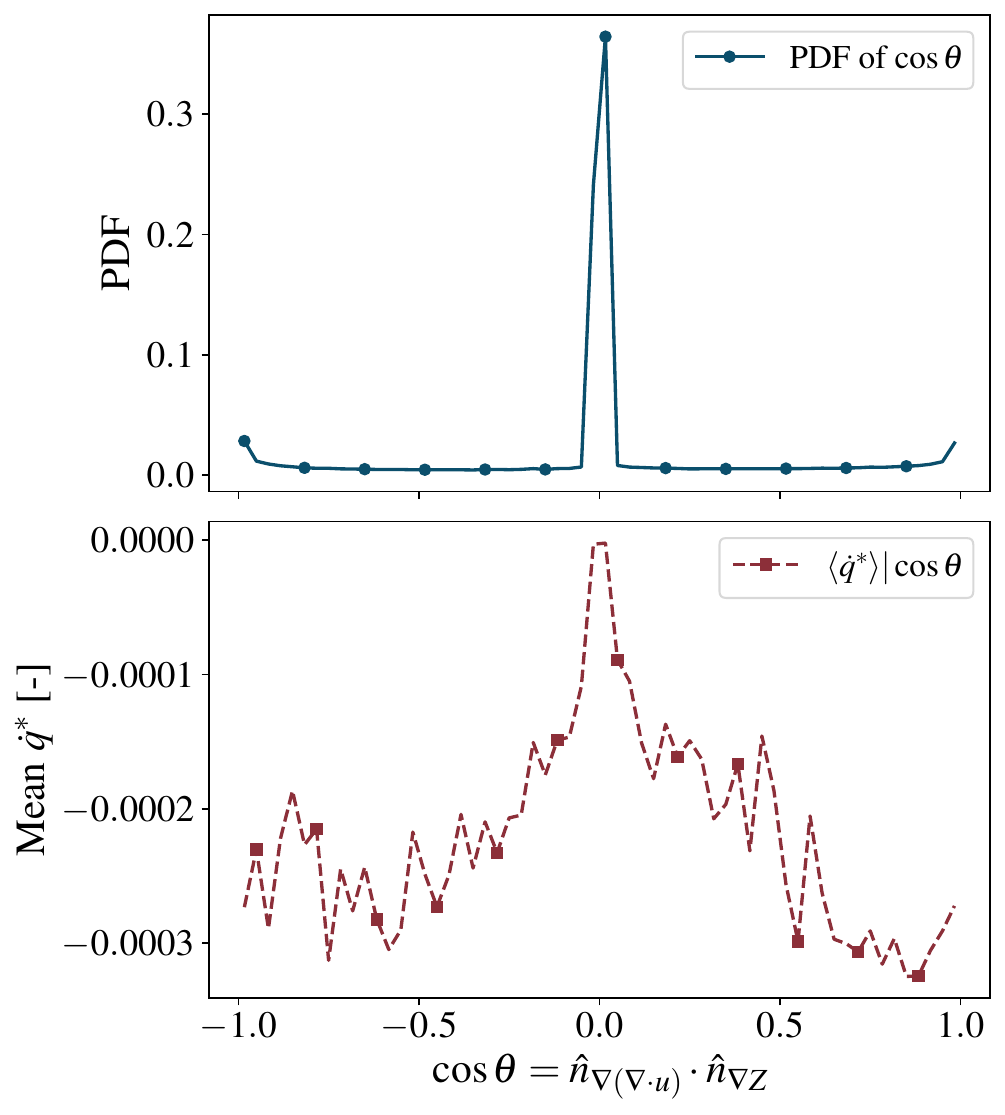}
\caption{Alignment between compression gradients and mixture gradients, together with the conditional mean $\dot{q}^*$ by alignment bin.}
  \label{fig:alignment}
\end{figure}

Event shape and field orientation add structural context. For an event component $\mathcal{C}_j$ with area $A_j$ and perimeter $P_j$, the compactness is
\begin{equation}
\Gamma_j=\frac{4\pi A_j}{P_j^2},
\end{equation}
and the eccentricity is computed from the covariance-tensor eigenvalues $\lambda_1\ge\lambda_2$ as
\begin{equation}
\varepsilon_j=\sqrt{1-\frac{\lambda_2}{\lambda_1}}.
\end{equation}
\noindent
These region-shape measures follow standard image-based morphology descriptors \cite{HaralickShapiro1992}.
\begin{figure}[t]
  \centering
  \includegraphics[width=\columnwidth]{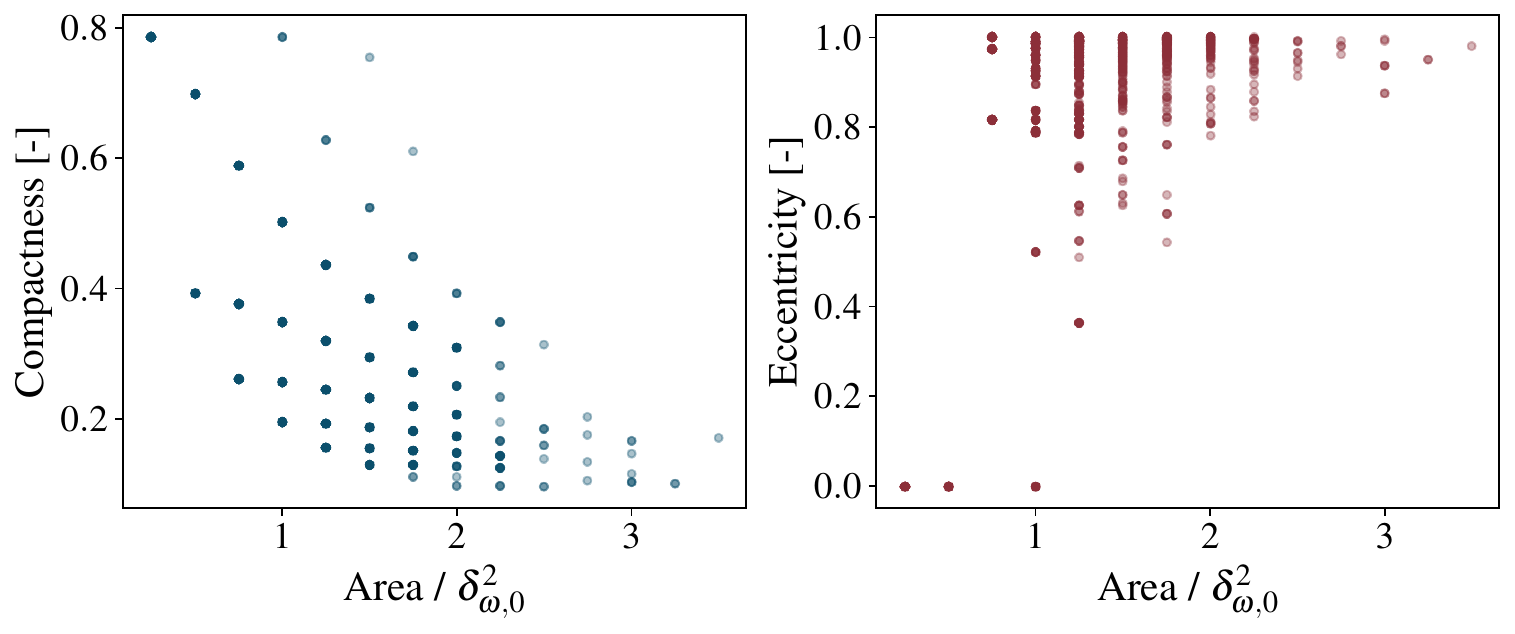}
\caption{Morphology metrics of compression regions using event area $A^*$, compactness, and eccentricity.}\label{fig:morphology}
\end{figure}
We also compute a boundary-curvature measure from the signed-distance function $\phi_s$ to the compression mask. With
\begin{equation}
\kappa = \nabla\cdot\left(\frac{\nabla\phi_s}{|\nabla\phi_s|}\right),
\end{equation}
the reported curvature is the mean of $|\kappa|$ over a narrow band around the compression boundary, scaled by $\delta_{\omega,0}$. This measure characterizes how wrinkled or fragmented the compression boundary is, while compactness and eccentricity characterize the event interiors. It is included as a secondary shape descriptor because curvature can distinguish a small number of smooth events from a population of more convoluted interfaces with similar total area.
\begin{figure*}[t]
  \centering
  \includegraphics[width=0.98\textwidth]{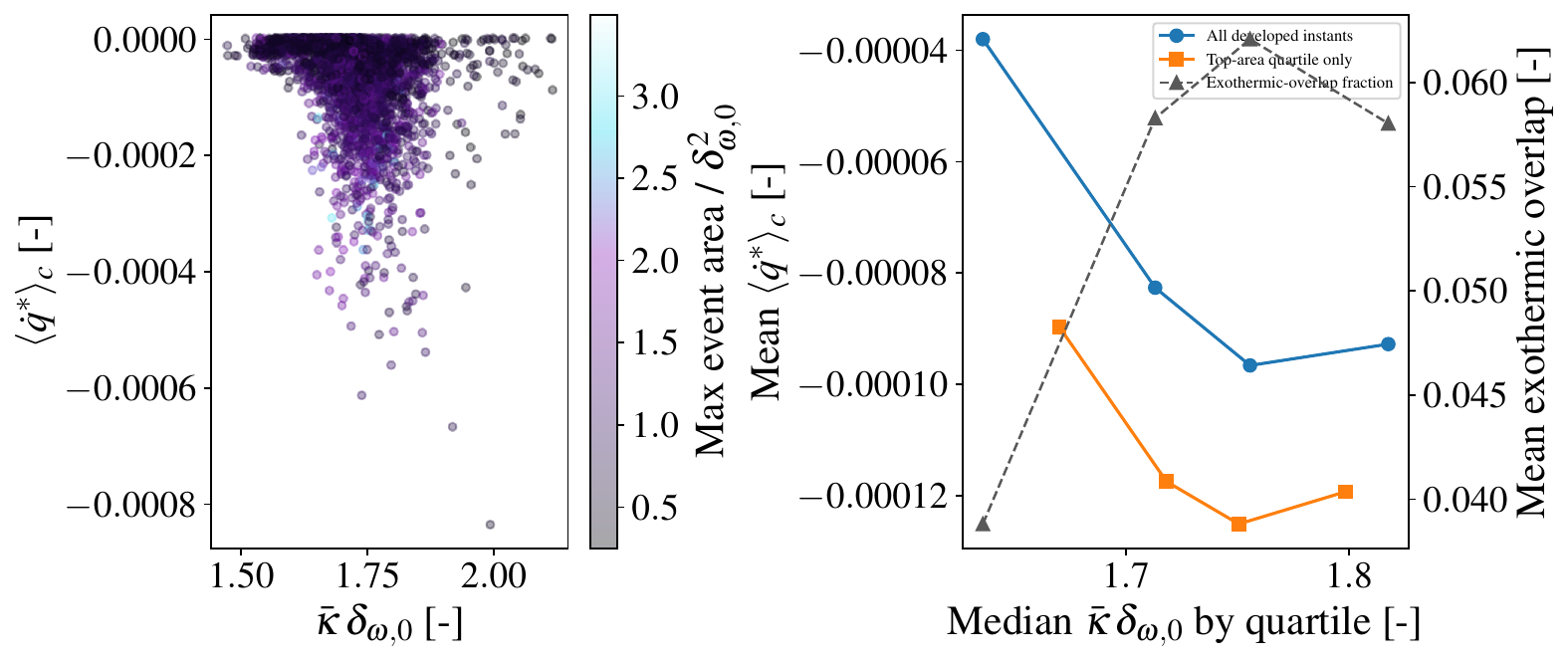}
\caption{Developed-regime curvature response. The scatter plot relates mean compression-boundary curvature to compression-conditioned heat release, with color denoting maximum event area. The quartile curves compare all developed instants with the top-area subset and show the accompanying exothermic overlap fraction.}
  \label{fig:curvature_followup}
\end{figure*}
Alignment is measured by
\begin{equation}
\cos\theta = \frac{\nabla(\nabla\cdot\bm{u})\cdot\nabla Z}
{|\nabla(\nabla\cdot\bm{u})|\,|\nabla Z|},
\end{equation}
\noindent
which is analogous to the orientation statistics used to study scalar-gradient alignment in turbulence \cite{AshurstEtAl1987}, and chemistry is conditioned on alignment bins to test whether exothermic response prefers specific relative orientations. The morphology panel in figure \ref{fig:morphology} uses physically normalized area, compactness, and eccentricity. Most compression objects are small and close to the small-scale compact end of the distribution, while the largest events become more elongated. Alignment statistics in figure \ref{fig:alignment} show that compression-gradient and mixture-gradient directions cluster near weak alignment on average. The most exothermic bins occur away from the neutral center, which suggests that alignment contains some chemistry information, with a weaker role than overlap and distance. These measures help characterize the structure of the compression population while preserving the main ranking established by the overlap and proximity results.

Figure \ref{fig:curvature_followup} shows how the boundary-curvature measure fits into that ranking. The strongest exothermic samples are concentrated in a subset of the developed record across a limited curvature range. Higher maximum-event area colors identify many of the most exothermic points, and the quartile curves show that curvature becomes most informative when interpreted together with event area and overlap. The top-area subset has a consistently stronger exothermic response than the all-instant curve at comparable curvature levels. The exothermic-overlap curve also rises across the lower-to-middle curvature quartiles, indicating that boundary complexity can accompany closer contact between compression and heat-release regions. At the highest curvature levels, the response saturates. Curvature refines the event-geometry picture. The dominant chemistry ordering still comes from overlap, proximity, and population measures.

\subsection{Synthesis}

The results point to a simple physical sequence. During startup, the fixed compression threshold selects sparse regions that carry little reactive information. During transition, compression begins to intersect the scalar layer, and high $|\nabla Z|^{2*}$ is the first chemistry-related quantity to respond strongly. Once the layer is developed, compression is organized into many finite-area events distributed across the shear layer. The exothermic response then depends on how those events overlap, or nearly overlap, the most reactive subset of the flow. Compression emerges first as a kinematic event population, sharpens scalar gradients next, and couples to heat release through a narrower reactive subset of the layer.

This ordering is physically plausible for a reacting shear layer. The mixture-fraction-gradient field is a kinematic scalar-mixing measure, so it can respond promptly when compression and strain sharpen scalar interfaces. Heat release is more constrained. Exothermic activity requires the appropriate mixture state and chemical progress in addition to local compression, and its strongest values occupy a narrower subset of the scalar layer. This explains why compression overlaps high $|\nabla Z|^{2*}$ more readily than it overlaps the most exothermic $\dot{q}$ regions, and why proximity to the exothermic set remains informative even when direct overlap is modest. The compression field organizes the conditions under which heat release is expressed, and the exothermic response remains spatially selective.

The temporal lag also supports this interpretation. The $|\nabla Z|^{2*}$ response peaks near the compression-area excursion, consistent with prompt scalar-gradient amplification in and around the compressive event population. The strongest exothermic response appears earlier than the maximum compression area, indicating that heat-release intensity is linked to the stage at which compression encounters already favorable reactive mixture. As compression continues to broaden or fragment, the area signal can increase after the most intense exothermic subset has already been sampled. Event-triggered timing, overlap, and distance capture that timing more clearly than the instantaneous compression footprint alone.

Shape descriptors, particularly compactness, eccentricity, curvature, and gradient alignment describe how compression is arranged, whereas overlap and distance indicate whether the event population lies on the reactive portion of the scalar layer. Large, elongated, or more curved events can create more opportunities for contact with reactive scalar interfaces. The chemistry ranking is strongest when that geometry is accompanied by overlap and proximity. The event description thus connects the global evolution of the temporal mixing layer to the local placement of compression, scalar gradients, and exothermic heat release.

\begin{table}[t]
\centering
\scriptsize
\setlength{\tabcolsep}{1.5pt}
\caption{Planar representativeness check using matched developed-regime three-dimensional fields. The mid-plane values are compared with the spanwise-plane distribution from the coarsened volume.}
\label{tab:planar_volume}
\begin{tabular}{@{}lccc@{}}
\toprule
Quantity & Mid-plane & Span. median & Span. range \\
\midrule
Compression support & 0.0483 & 0.0499 & [0.0472, 0.0521] \\
$10^4\langle\dot{q}^*\rangle_c$ & -2 & -2.09 & [-2.39, -1.73] \\
$\chi_q$ & 0.183 & 0.183 & [0.155, 0.21] \\
$\chi_g$ & 0.158 & 0.164 & [0.149, 0.198] \\
\bottomrule
\end{tabular}
\end{table}

Table \ref{tab:planar_volume} compares the mid-plane record with matched three-dimensional fields. For the event quantities used in the paper, the mid-plane values lie within the spanwise range for compression support, compression-conditioned heat release, and both overlap measures. The agreement is strongest for the quantities that control the main physical interpretation: compression support and compression--chemistry overlap. The mid-plane record acts as a representative section for these event-level statistics, while the interpretation remains planar in scope.

\noindent
The resulting hierarchy is consistent across the different views of the data. Compression--chemistry coupling is clearest when compression is treated as a time-dependent event population and the chemistry response is read through overlap, proximity, and lag. Startup and transition show how the coupling emerges; the developed regime supplies the stable statistics needed to order the measures. Within that regime, scalar-gradient amplification is the more immediate signature of compression, and the exothermic response is more selective in space and offset in time. Event geometry and alignment add structural context to the dominant picture given by event organization, overlap, and distance.

\FloatBarrier
\section{Conclusions}
This study used a fixed-threshold compression-event description to examine how compression organizes scalar-gradient amplification and exothermic chemistry in a reacting supersonic temporal mixing layer. The temporal record separates into startup, transition, and developed regimes. That separation is important because the compression mask carries little reactive information early in the record, then becomes a statistically meaningful event population once the layer reaches the developed regime.

In the developed regime, compression is distributed across the shear layer as a population of intermittent connected events. Stronger exothermic response is associated with larger maximum-event area, larger event count, larger compression--chemistry overlap, and smaller distance to exothermic heat-release regions. The timing also separates the two chemistry-related responses. The conditioned $|\nabla Z|^{2*}$ response peaks near zero lag, consistent with prompt scalar-gradient amplification, while the largest-magnitude exothermic response precedes peak compression-area excursions by approximately $\Delta t^*\approx -0.85$. Compression is most informative when its event organization and its placement relative to the reactive subset of the layer are retained.

Event morphology and gradient alignment add structural detail to this picture. They describe how compression is arranged, while overlap and distance more directly rank the reactive response. For combustion modeling and reduced descriptions, this distinction suggests that compression-sensitive indicators should retain event population and compression-to-reaction placement, since local dilatation alone does not identify the strongest exothermic subset of the reacting layer.

The sampled mid-plane record defines the scope of the interpretation. The analysis resolves event-level organization, relative timing, and spatial placement within that plane, with the matched-volume check indicating that the mid-plane lies within the spanwise distribution for the main event quantities. A corresponding volumetric extension would extract compression regions and reactive interfaces throughout the full field and quantify how planar near-overlap events intersect connected three-dimensional structures. The quantities emphasized here--fixed dilatation-based compression support, event population, compression-to-chemistry overlap, compression-to-chemistry distance, and lag relative to compression-area excursions--provide a compact basis for comparing additional reacting shear-layer datasets.

\section*{Acknowledgments}
The computing power for this study was provided by the Phoenix Computing Cluster as a part of Georgia Tech's Partnership for Advanced Computing Environment, and is gratefully acknowledged.

\section*{Funding}
This research received no specific grant from any funding agency, commercial or not-for-profit sectors.

\section*{Declaration of competing interest}
The authors report no conflict of interest.

\section*{Data availability}
The data that support the findings of this study are available from the corresponding author upon reasonable request.

\section*{CRediT authorship contribution statement}
Sriram P. Kalathoor: Conceptualization, Methodology, Software, Formal analysis, Investigation, Visualization, Writing -- original draft. Joseph C. Oefelein: Conceptualization, Resources, Supervision, Writing -- review and editing.

\section*{Declaration of generative AI and AI-assisted technologies in the manuscript preparation process}
\textbf{Placeholder:} Gemini was used to generate python code snippets for the colormaps used in Fig.~\ref{fig:structure}, and to generate hex codes for the red/blue colors used in all the lineplots.

\bibliographystyle{cnf-num}
\bibliography{cnf-refs}

\end{document}